\documentstyle[11pt,aaspp4]{article}
\twocolumn
\lefthead{Woolf \&\ Lambert}
\righthead{Young HgMn star in Orion}

\begin{document}

\title{Three very young HgMn stars in the Orion~OB1 Association}

\author{Vincent M. Woolf and  David L. Lambert}
\affil{Department of Astronomy and McDonald Observatory, University of Texas,
\\Austin, Texas 78712
\\Electronic mail: vincent@shaka.as.utexas.edu, dll@astro.as.utexas.edu}

\begin{abstract}
We report the detection of three mercury-manganese stars in the Orion~OB1
association.
HD~37886 and BD--$0^\circ$984 are in the approximately 1.7~million year old
Orion~OB1b. HD~37492 is in the approximately 4.6~million year old Orion~OB1c.
Orion~OB1b is now the youngest cluster with known HgMn star members.  This
places an observational upper limit on the time scale needed to produce the
chemical peculiarities seen in mercury-manganese stars, which should help in
the search for the cause or causes of the peculiar abundances in HgMn and other
chemically peculiar upper main sequence stars.
\end{abstract}


\section{Introduction}
Mercury-manganese (HgMn) stars are a class of chemically peculiar (CP) upper
main sequence stars which show bizarre atomic and isotopic abundance patterns
in their spectra.  They tend to show large overabundances of Mn, Pt, Au, Hg, and
several other elements.  Many are helium-weak.  One of the striking
peculiarities found in HgMn stars is the presence of unusual isotopic abundance
mixtures for Hg, Pt, and Tl which cannot be explained by any known nuclear
processes. The origins of the chemical peculiarities in HgMn and other
CP stars are still an open question.

Until quite recently, it was generally accepted that radiation pressure driven
diffusion was causing the peculiar atomic and isotopic abundances observed
in HgMn stars: radiation pressure levitates Hg, Mn, etc. ions in the atmosphere,
concentrating them in the line-forming region of stars which began with
``normal'' abundances (Michaud 1970, Michaud, Reeves,
\&\ Charland 1974).  Some recent works have found that it is difficult to
explain the Hg atomic and isotopic abundances using diffusion alone.
Proffitt et~al. (1999) report that at no level in the photosphere of
$\chi$~Lupi, a bright HgMn star observed in detail with the  HST,
is the radiation force on Hg greater than the gravitational force, making it
impossible for diffusion in the photosphere to create the observed
Hg overabundance.  Woolf \&\ Lambert (1999) found Hg
isotopic mixtures in two HgMn stars, 11~Per and HR~7245, which are very
difficult, if not impossible, to explain using diffusion alone.  These recent
developments make it important to explore other processes which may contribute
to creating the peculiar abundances in HgMn stars.  These include light induced
drift, diffusion below the photosphere, disorganized magnetic fields, and winds.

Stellar parameters which may influence the mechanisms which create peculiar 
abundances include temperature, surface gravity, initial chemical
composition, stellar rotation, magnetic fields, presence of a binary companion,
and stellar age.  Exploration of this multi-parameter space should yield clues
to the origins of CP stars in general and HgMn stars in particular.

As part of an exploration of HgMn stars and stellar ages,
we have observed late B and early A stars in the Orion~OB1 association.
The Orion~OB1 association contains
subgroups which are among the youngest spectroscopically observable clusters.
The detection of very young CP stars puts an observational upper limit on the
length of time needed to produce their chemical peculiarities.

\section{Observations and analysis}

Stars were selected for observation from known Orion~OB1 stars listed in Warren
\&\ Hesser (1978) and were found using using SIMBAD coordinates and, when
necessary, the star charts of Warren \&\ Hesser (1977) and Parenago (1954).

Observations were made at McDonald Observatory 1998 December 24--28 and
1999 March 5--11 using the
cross dispersed echelle spectrometer at the coud\'e focus of the 2.7~m
telescope (Tull et~al.\ 1995).  The resolving power was
approximately 60,000.  Since
the spectra were to be used to search for CP stars, not high precision
abundance analyses, very high signal-to-noise spectra were not a priority,
though stars were re-observed to provide higher signal if interesting spectral
features were detected.  Most spectra had signal-to-noise ratios between
70 and 100.  At these low signal-to-noise levels we may not
detect some CP stars with broad lines and/or only weakly enhanced abundances.
Spectra were reduced using standard IRAF procedures for bias and scattered
light subtraction, flat lamp division, spectrum reduction, and ThAr wavelength
calibration.

HD~37492 and BD--$0^\circ$984 were
found to have detectable 3984~\AA\ Hg~II lines.  BD--$0^\circ$984
and HD~37886 were
found to have Mn~II lines which were obviously enhanced on
visual inspection.  The signal-to-noise levels in spectra used for abundance
analysis for these three stars varies with wavelength.  In the spectral order
containing the 3984~\AA\ Hg~II line signal-to-noise is about 70, 200, and
140 in HD~37886, BD--$0^\circ$984, and HD~37492, respectively.

Mn abundances were calculated from equivalent widths of Mn~II lines using
MOOG (Sneden 1973).  For HD~37886, HD~37492, and BD--$0^\circ$984, 31, 18,
and 43 lines were used, respectively, with $gf$ values and wavelengths
obtained from the VALD database (Piskunov et~al. 1995)
in which most Mn~II data are
originally taken from Kurucz (1994).  No corrections were made for desaturation
due to hyperfine splitting of Mn~II lines in this analysis.  This may cause
us to overestimate Mn abundances (Jomaron, Dworetsky, \&\ Allen 1999).
However, calculating $A$(Mn) for BD--$0^\circ$984 using only the 12 lines with
equivalent widths smaller than 35~m\AA, which should not be strongly
affected by hyperfine desaturation, gives the same abundance as a calculation
including the stronger lines.  Corrections
made to Mn abundances due to hyperfine splitting and possible line
blending would still leave Mn enhanced in all three of these stars.

Hg abundances were estimated by fitting synthetic
spectra to the 3984~\AA\ Hg~II line using MOOG as described in Woolf \&\ 
Lambert (1999).  Calculations were made using individual isotopic and
hyperfine components and varying Hg isotopic abundances independently of
each other.  The reported Hg abundance is the sum of the isotopic abundances.

The model atmosphere used
for the estimates was calculated with ATLAS9 (Kurucz 1993).
T$_{\rm eff}$ and log$g$ were estimated using Stromgren
{\it uvby}$\beta$ photometry (Mermilliod, Mermilliod, \&\ Hauck 1997) and a
FORTRAN program (Moon 1985, Napiwotzki, Sch\"onberner, \&\ Wenske 1993)
which interpolates the grids of Moon \&\ Dworetsky (1985).  Microturbulences
were estimated by requiring that there be no slope in $A$(Fe~II) vs
equivalent width for each star.

In the 1999 March 10 spectrum of HD~37492 most of
the strong lines, though not the Mn~II lines, have a weaker line just
to the blue, indicating that HD~37492 is a double-lined binary with a HgMn
primary and a non-HgMn secondary.  The presence of a binary companion was not
taken into account when using photometry to calculate stellar parameters or
calculating abundances using equivalent width or spectral synthesis analysis.
The actual effective temperature of the primary
is probably hotter than that used and the
equivalent widths of the lines would be larger if a correction was made for the
companion's continuum flux contribution.

\section{Results and discussion}

\begin{figure}
\plotfiddle{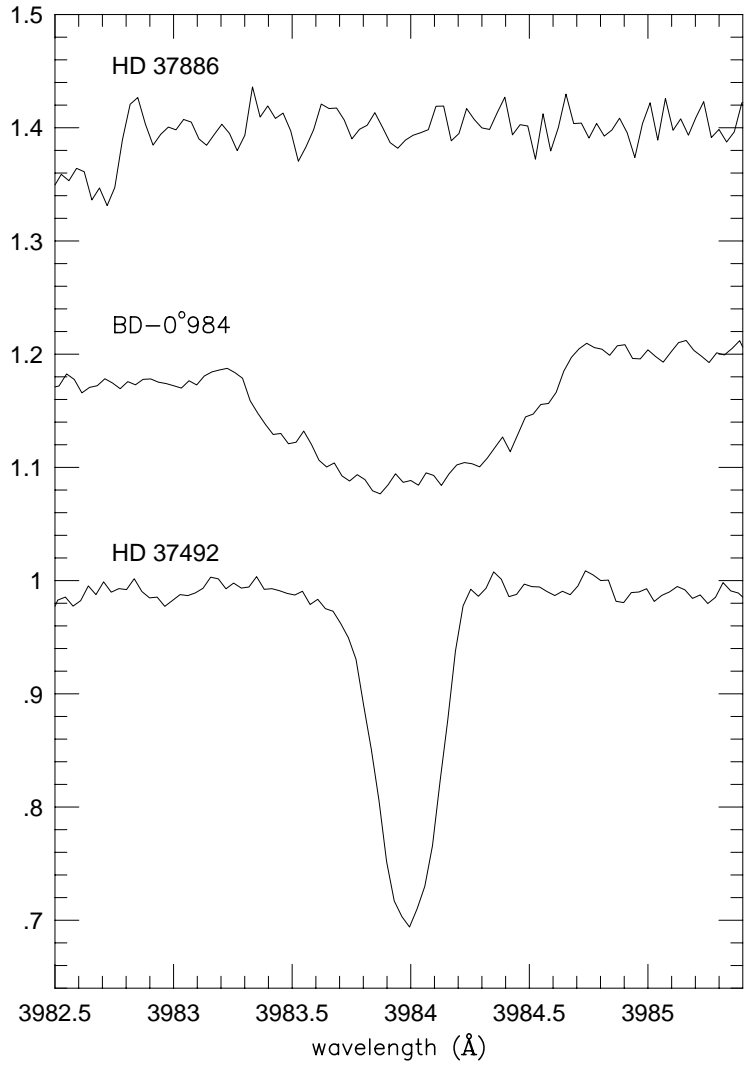}{4in}{0}{60}{60}{-90}{0}
\caption{3984~\AA\ Hg~II line regions of HD~37886, BD--$0^\circ$984,
and HD~37492.}
\label{f1}
\end{figure}

Figure~\ref{f1} shows the spectral region containing the 3984~\AA\ Hg~II
line in the HgMn stars HD~37886,
BD--$0^\circ$984, and HD~37492. Table~1 gives the
values for T$_{\rm eff}$, log$g$, and $\xi$
used in the abundance determinations;
the $v$sin$i$ values; Fe, Mn, and Hg abundances; and central wavelengths and
equivalent widths of the 3984~\AA\ Hg~II line found for the three stars.
The central wavelengths of the Hg~II lines of BD--$0^\circ$984 and HD~37492
imply that they have Hg isotopic compositions similar to those of the
sharp-lined HgMn stars HR~1800,
$\iota$~CrB, and HR~4072 (Woolf \&\ Lambert 1999): their Hg mixtures are
weighted toward the heavier isotopes, as are the mixtures of the large majority
of HgMn stars, though not as strongly as in $\chi$~Lupi or other stars which
show pure $^{204}$Hg.

Hg is very overabundant in HD~37492 and BD--$0^\circ$984.
The 3984~\AA\ Hg~II line
is undetectable in HD~37886.  The Hg abundances in HgMn
stars range from undetectable to about  $A({\rm Hg}) \approx 7.0$ or 680,000
times the solar system abundance\footnote{We use the
notation: $A(X) \equiv log_{10}(N_{\rm X}/N_{\rm H}) + 12.00$.}
(Smith 1997).  Mn is overabundant in
HD~37886, BD--$0^\circ$984, and HD~37492.
Mn abundances in HgMn stars have been previously reported to be as large as
$A({\rm Mn}) \approx 8.0$ or about 400 times the solar abundance
(Smith \&\ Dworetsky 1993, Jomaron et~al. 1999).
The Mn abundance found for
BD--$0^\circ$984, about 800 times solar, is large even for
a HgMn star, though its large uncertainty would allow its Mn abundance to be
as low as 300 times solar.
The Hg and Mn abundances found for these three stars show that they are
HgMn stars.

The lack of a detectable Hg line in HD~37886 is unusual for a HgMn star, but
not unique.  For example, 53~Tau, HR~2676, and 30~Cap also have no detectable
3984~\AA\ line (Smith 1997, Woolf \&\ Lambert 1999).  HD~37886 and other
HgMn stars not strongly enhanced in Hg do not have temperatures,
gravities, or microturbulences which would distinguish them from other HgMn
stars.  An in-depth study of known low-Hg HgMn stars may suggest clues to
the mechanisms creating CP stars.

Two stars out of 38 B8 to A2 stars observed in Orion~OB1b were found to be
HgMn stars.  One star out of 23 B8 to A2 stars observed in Orion~OB1c was
found to be a HgMn star. No HgMn stars were found in Orion~OB1d, the
Trapezium Cluster, but only 11 B and A stars were observed in OB1d, so the
non-detection does not necessarily put a lower limit on the time scale to
create a HgMn star.
No stars other than the three HgMn stars discovered had detectable
3984~\AA\ Hg~II lines or Mn~II lines in the 3800~\AA\ to 9900~\AA\ region.

Detecting a HgMn star depends on how enhanced the Hg
and/or Mn lines are, the star's $v$sin$i$, the noise level of the spectrum,
and the wavelength region observed.  We cannot rule out the possibility that
one or more of the stars we observed
is a HgMn star but that we failed to detect it because it is only weakly
enhanced in Hg and Mn and has a high $v$sin$i$: the higher the $v$sin$i$, the
stronger the line has to be to be detected.  For example, in stars with
$v\sin i \lesssim 10~{\rm km/s}$ we can rule out Hg abundances greater than
$A{\rm (Hg)} > 4.0$, but for stars with $v\sin i > 100~{\rm km/s}$ and signal
to noise of about 80 we would not detect the Hg~II line for
$A{\rm (Hg)} \lesssim 5.2$,
depending somewhat on temperature, gravity, and isotopic mix.
Twenty of the 38 B8 to A2 stars
observed in Orion~OB1b and 11 of the 23 stars observed in Orion~OB1c had
$v\sin i \lesssim 100~{\rm km/s}$, the limit normally quoted for HgMn stars.
The number of HgMn stars found in our search is not inconsistent with
what would be
expected in field stars.  Approximately 5\% of A0 stars and 12\% of B9 stars
are HgMn stars (Smith 1996).

The Orion~OB1 association is divided into four
subgroups: 1a, 1b, 1c, and 1d.  The groups are $11.4 \pm 1.9$, $1.7 \pm 1.1$,
$4.6^{+1.8}_{-2.1}$, and $<1.0$~Myr old, respectively (Brown, de~Geus, \&\
Zeeuw 1994).  HD~37492 in the Orion~OB1c subgroup is presumably about
4.6~million years old.  HD~37886 and BD--$0^\circ$984 are in the Orion~OB1b
subgroup and are probably between 0.6 and 2.8~million years old.
The age determinations of Brown et~al. differ from previous reports
which found OB1b, at about 5~Myrs, older than OB1c, at about 4~Myrs.
(Warren \&\ Hesser 1978).
Brown et~al. (1994) estimated ages using theoretical isochrones.
Warren \&\ Hesser (1978) estimated ages using the bluest main sequence stars.

HgMn stars have been found in other young clusters.
The HgMn stars HR~1690 and HR~1800 are in the 11.4~million year old Orion~OB1a.
The HgMn star HR~5998 is in the 5 to 8~million year old (de~Geus,
de~Zeeuw, \&\ Lub 1989, de~Zeeuw \&\ Brand 1985) Upper Scorpius subgroup of the
Scorpius-Centaurus OB Association.  The HgMn star HD~47553 is in the 6.5~million
year old NGC~2264 (Malysheva 1997, Pyatkes 1993).  The ages determined for
these young clusters are still fairly uncertain and there is apparently an age
spread among the stars in each cluster.  HD~37886 and BD--$0^\circ$984 in
Orion~OB1b are the youngest known HgMn stars in a cluster of a known age.

The time scales for the various processes which have been proposed as causes
of the chemical peculiarities in HgMn stars are not very well known.  Michaud
(1970) calculated a radiative diffusion time scale of about $10^4$ years.
More recently, Seaton (1998, 1999) calculated that for some Fe peak elements,
such as Mn and Ni, the time for diffusion to create maximum concentrations
is increased to about $4 \times 10^6$~years by the need for atoms to pass
through a region where ions have Ar-like electronic configurations
which have few strong absorption lines in the wavelength region where the
stellar flux is strong.
However, as discussed in the introduction, calculations by Proffitt et~al.
(1999) have shown that the model of Michaud (1970) in which Hg is concentrated
by radiative diffusion alone will not work in the HgMn star $\chi$~Lupi.
We know of no calculations of time scales for light induced drift, diffusion
below the photosphere, or other proposed element concentrating processes.
Diffusion, light induced drift, and other
processes which require a quiescent atmosphere to operate would only have an
effect after the He~II convection zone disappears in a young star as He
diffuses out of the bottom of the zone.  It is possible that the time scale
for the He~II convection zone to disappear, on the order of $10^6$ years
(Vauclair, Vauclair, \&\ Pamjatnikh 1974), is the major limiting factor on how
quickly HgMn stars can be produced.

The three very young HgMn stars in Orion~OB1b and OB1c show different abundance
patterns.
HD~37886 is enhanced in Mn but not in Hg.  HD~37492 is fairly sharp-lined
and strongly enhanced in Hg but only weakly in Mn.  BD--$0^\circ$984
is broad-lined
and is strongly enhanced in both Hg and Mn.  Its spectrum is almost identical
in both line strengths and line broadening 
to that of $\alpha$~And, the first HgMn star identified.
The HgMn stars found in Orion provide constraints for models developed to
explain HgMn star peculiarities.  Their youth means that the peculiarities
develop quickly, and possibly in the pre-main sequence phase. Pre-main sequence
chemical peculiarities could be produced as a result of fractionation of
accretion from a circumstellar shell or disk.
The differences in the Hg and Mn abundances in stars in
the same association mean that age and
initial chemical composition do not alone determine the abundance patterns
seen in HgMn stars.

\acknowledgments

This research has made use of the Simbad database, operated at CDS, Strasbourg,
France and the Vienna atomic line database (VALD), operated at Institut
f\"ur Astronomie, Vienna,
Austria.  The authors acknowledge the support of the National Science
Foundation (grant AST 96-18414) and the Robert A. Welch Foundation of Houston,
Texas.

\pagebreak

\clearpage

\begin{deluxetable} {llll}
\normalsize
\tablecaption{Parameters and abundances for Orion HgMn stars}
\tablewidth{290pt}
\tablehead{
& \colhead{HD~37886} & \colhead{BD$-0^\circ 984$} &
\colhead{HD~37492\tablenotemark{a}}
}
\startdata
Orion subgroup          & OB1b          & OB1b          & OB1c\cr
T$_{\rm eff} (K)$       & 12540         & 13975         &13325\cr
log$g$                  & 4.20          & 4.20          &3.93\cr
$\xi$ (km/s)            & 1.0           & 2.0           &2.0\cr
$v$sin$i$ (km/s)        &$19 \pm 1$     &$49 \pm 2$     &$8.5 \pm 1.0$\cr
$A$(Fe~II)\tablenotemark{b}             &$7.40 \pm 0.22$&$7.85 \pm 0.35$&$7.91 \pm 0.21$\cr
$A$(Mn~II)\tablenotemark{b}             &$7.64 \pm 0.35$&$8.30 \pm 0.40$&$6.63 \pm 0.20$\cr
$A$(Hg~II)      \tablenotemark{b}       &$<4.50$        &$5.88 \pm 0.29$&$5.60 \pm 0.15$\cr
Hg central $\lambda$\ (\AA)     &\nodata & 3983.971 &3983.981 \cr
3984~\AA\ $\rm W_\lambda$ (m\AA)&\nodata &115   &96     \cr
\enddata
\tablenotetext{a}{HD~37492 data ignores binary companion flux contribution
both to photometry used for estimating stellar parameters and to spectra used
for abundance calculations.}
\tablenotetext{b}{Solar abundances are $A{\rm (Fe)} = 7.50$,
$A{\rm (Mn)} = 5.39$, and $A{\rm (Hg)} = 1.17$ (Grevesse, Noels, \&\ Sauval
1996).}
\end{deluxetable}


\begin{thebibliography}{}
\bibitem {bdz94} Brown, A.~G.~A., de~Geus, E.~J., \&\ de~Zeeuw, P.~T. 1994,
A\&A, 289, 101
\bibitem {ddl89} de~Geus, E.~J., de~Zeeuw, P.~T., \&\ Lub, J. 1989, A\&A, 216,
44
\bibitem {db85} de~Zeeuw, P.~T., \&\ Brand, J. 1985, in Birth and Evolution of
massive stars and stellar groups, eds. H. van~Woerden, \&\ W. Boland, Ap\&SS
Library, Vol 120, 95
\bibitem {gns96} Grevesse, N., Noels, A., \&\ Sauval, A.~J. 1996, in Cosmic
Abundances, ASP Conf. Ser. 99, ed. S.~S. Hold \&\ G. Sonnebrrn (San Francisco:
ASP), 117
\bibitem {jda99} Jomaron, C.~M., Dworetsky, M.~M., \&\  Allen, C.~S. 1999,
MNRAS, 303, 555
\bibitem {k93} Kurucz, R.~L. 1993, in Peculiar Versus Normal Phenomena in
A-Type and Related Stars, ASP Conf. Ser. 44, ed. M.~M. Dworetsky, F. Castelli
\&\ R. Faraggiana (San Francisco: ASP), 87
\bibitem {k94} Kurucz R.~L., 1994, SAO, Cambridge, CDROM 20--22
\bibitem {m97} Malysheva, L.~K. 1997, Astron. Let., 23, 585
\bibitem {mmh97} Mermilliod, J.~-C., Mermilliod, M., \&\ Hauck, B. 1997, A\&AS,
124, 349
\bibitem {m70} Michaud, G. 1970, ApJ, 160, 641
\bibitem {mmc81} Michaud, G., Reeves, H., \&\ Charland, Y. 1974, A\&A, 37, 313
\bibitem {mo85} Moon, T.~T. 1985, Comm. Univ. London Obs. 78, 1
\bibitem {md85} Moon, T.~T., \&\ Dworetsky, M.~M. 1985, MNRAS, 217, 305
\bibitem {nsw93} Napiwotzki, R., Sch\"onberner, D., \&\ Wenske, V. 1993, A\&A,
268, 653
\bibitem {p54} Parenago, P.~P. 1954, Trudy Gosud. Astron. Inst. Shternberga,
25, 1
\bibitem {pk95} Piskunov N.~E., Kupka F., Ryabchikova T.~A., Weiss W.~W.,
Jeffery C.~S. 1995, A\&AS 112, 525
\bibitem {pr99} Proffitt, C.~R., Brage, T., Leckrone, D.~S., Wahlgren, G.~M.,
Brandt, J.~C., Sansonetti, C.~J., Reader, J., \&\ Johansson, S.~G. 1999, ApJ, 
512, 942
\bibitem {p93} Pyatkes, S.~A. 1993, Bull. Special Astrophys. Obs., 35, 76
\bibitem {se98} Seaton, M.~J. 1998, in Highlights of Astronomy, Vol. 11B,
ed. J.~Andersen (Dordrecht, Netherlands: Kluwer), 664
\bibitem {se99} Seaton, M.~J. 1999, MNRAS, in press
\bibitem {sm96} Smith, K.~C. 1996, Ap\&SS, 227, 77
\bibitem {sm97} Smith, K.~C. 1997, A\&A, 219, 928
\bibitem {sd93} Smith, K.~C. \&\ Dworetsky, M.~M. 1993, A\&A, 274, 335
\bibitem {s73} Sneden, C. 1973, Ph.D. thesis, University of Texas
\bibitem {tul95} Tull, T.~G., MacQueen, P., Sneden, C., \&\ Lambert, D.~L. 1995,
PASP, 107, 251
\bibitem {vvp74} Vauclair, G., Vauclair, S., \&\ Pamjatnikh, A. 1974, A\&A, 31,
63
\bibitem {wh77} Warren, W.~H., \&\ Hesser, J.~E. 1977, ApJS, 34, 115 
\bibitem {wh78} Warren, W.~H., \&\ Hesser, J.~E. 1978, ApJS, 36, 497
\bibitem {wl99} Woolf, V.~M., \&\ Lambert, D.~L. 1999, ApJ, 521, in press
\end{thebibliography}
\end{document}